\documentclass[12pt,preprint]{aastex}

\usepackage{emulateapj5}

\usepackage{amssymb,amsmath}
\newcommand{\xI}{\boldsymbol{\chi}_i}
\newcommand{\xa}{\boldsymbol{\chi}_1}
\newcommand{\xb}{\boldsymbol{\chi}_2}
\newcommand{\hLN}{\mathbf{\hat{L}}}
\newcommand{\LN}{\mathbf{L}}
\newcommand{\Si}{\mathbf{S}_i}
\newcommand{\Sa}{\mathbf{S}_1}
\newcommand{\Sb}{\mathbf{S}_2}
\newcommand{\qsum}{1+q}
\newcommand{\VD}{\boldsymbol{\Delta}}
\newcommand{\ez}{\mathbf{\hat{e}}_z}
\newcommand{\DP}{\Delta\phi}
\newcommand{\Oi}{\boldsymbol{\Omega}_i}

\newcommand{\beq}{\begin{equation}}
\newcommand{\eeq}{\end{equation}}
\newcommand{\ba}{\begin{eqnarray}}
\newcommand{\ea}{\end{eqnarray}}
\def\spose#1{\hbox to 0pt{#1\hss}}
\newcommand{\lta}{\mathrel{\spose{\lower 3pt\hbox{$\mathchar"218$}}
      \raise 2.0pt\hbox{$\mathchar"13C$}}}
\newcommand{\gta}{\mathrel{\spose{\lower 3pt\hbox{$\mathchar"218$}}
      \raise 2.0pt\hbox{$\mathchar"13E$}}}

\shorttitle{Relativistic Recoil Suppression}
\shortauthors{Kesden, Sperhake, \& Berti}

\begin{document}

\title{Relativistic Suppression of Black Hole Recoils}

\author{Michael Kesden\altaffilmark{1}, Ulrich Sperhake\altaffilmark{1,2}, and
Emanuele Berti\altaffilmark{1,2}}

\altaffiltext{1}{California Institute of Technology, MC 350-17, 1216
E. California Blvd., Pasadena, CA 91125}
\altaffiltext{2}{Department of Physics and Astronomy, The University of
  Mississippi, University, MS 38677-1848}%, berti@phy.olemiss.edu}

\begin{abstract}
Numerical-relativity simulations indicate that the black hole produced
in a binary merger can recoil with a velocity up to $v_{\rm max}
\simeq 4,000$ km/s with respect to the center of mass of the initial
binary.  This challenges the paradigm that most galaxies form through
hierarchical mergers, yet retain supermassive black holes at their
centers despite having escape velocities much less than $v_{\rm max}$.
Interaction with a circumbinary disk can align the binary black hole
spins with their orbital angular momentum, reducing the recoil
velocity of the final black hole produced in the subsequent merger.
However, the effectiveness of this alignment depends on highly
uncertain accretion flows near the binary black holes.  In this {\it
Letter}, we show that if the spin $\textbf{S}_1$ of the more massive
binary black hole is even partially aligned with the orbital angular
momentum $\textbf{L}$, relativistic spin precession on sub-parsec
scales can align the binary black hole spins with each other.  This
alignment significantly reduces the recoil velocity even in the
absence of gas.  For example, if the angle between $\textbf{S}_1$ and
$\textbf{L}$ at large separations is 10 degrees while the second spin
$\textbf{S}_2$ is isotropically distributed, the spin alignment
discussed in this paper reduces the median recoil from 864 km/s to 273
km/s for maximally spinning black holes with a mass ratio of 9/11.
This reduction will greatly increase the fraction of galaxies
retaining their supermassive black holes.
\end{abstract}


%\maketitle

%--------------------------------------------------------
\section{Introduction}
\label{S:intro}

Observations suggest that most galaxies host supermassive black holes
(SBHs) at their centers whose masses are tightly correlated with
properties of their host spheroids
\citep{Magorrian:1997hw,Ferrarese:2000se,Tremaine:2002js}.  If
galaxies form through hierarchical mergers, their SBHs may form from
the merger of the smaller SBHs in their progenitor galaxies.  The
final stage of these black-hole mergers involves highly curved,
dynamical spacetime that can only be simulated with fully nonlinear
numerical relativity (NR).  Following a major breakthrough in 2005
\citep{Pretorius:2005gq,Campanelli:2005dd,Baker:2005vv},
numerical relativists can now accurately determine the anisotropic
emission of gravitational waves during the final stage of black-hole
mergers.  When gravitational waves are preferentially emitted in one
direction during a merger, conservation of linear momentum requires
that the final black hole produced in that merger recoil in the
opposite direction.  These recoil velocities or ``kicks'' can approach
4,000 km/s for maximally spinning mergers
\citep{Campanelli:2007ew,Gonzalez:2007hi}.  Kicks this large exceed
the escape velocities of even the most massive galaxies, and would
thus eject SBHs from their hosts
\citep{Merritt:2004xa}.  Frequent SBH ejections would seem to
contradict the tightness of the observed correlations between SBHs and
their host galaxies.  Kicks would pose an even greater problem at high
redshifts, where typical galactic escape velocities decrease while
recoils remain a fixed fraction of the speed of light.

How might we avoid black-hole mergers that lead to large kicks?  To
answer this question, we must take a closer look at how the predicted
recoils depend on the dimensionless spins $\xI \equiv \Si/m_{i}^2$ and
mass ratio $q \equiv m_2/m_1 \leq 1$ of the merging black holes.
Reliable NR simulations have been performed for $q \geq 0.1$ and
$|\xI| \leq 0.9$; in this range the recoils are well described by the
fitting formula \citep{Campanelli:2007ew}
\beq \label{E:ktot}
\vec{v}(q,\xa,\xb) = v_m \mathbf{\hat{e}}_1 + v_\perp
(\cos \xi~\mathbf{\hat{e}}_1 + \sin \xi~\mathbf{\hat{e}}_2)
+ v_\parallel \ez
\eeq
where
\begin{eqnarray} \label{E:vm}
	v_m &=& A\eta^2 \frac{1-q}{\qsum}(1 + B\eta)~, \\ \label{E:vper}
	v_\perp &=& H\eta^2 \VD^\parallel \cdot \ez~, \\ \label{E:vpar}
	v_\parallel &=& K\eta^2 \cos(\Theta - \Theta_0) |\VD^\perp|~.
\end{eqnarray}
Here ($\mathbf{\hat{e}}_1, \mathbf{\hat{e}}_2, \ez$) are an orthonormal basis
with $\ez$ parallel to the orbital angular momentum $\LN$, $\eta \equiv
q/(\qsum)^2 \leq 1/4$ is the symmetric mass ratio, and $\VD^{\parallel,\perp}$
are the components of 
\beq \label{E:Delta} 
\VD \equiv \frac{q\xb - \xa}{\qsum} 
\eeq
parallel and perpendicular to $\LN$.  $\Theta$ is the angle between
$\VD^\perp$ and the separation $\mathbf{r}$ of the two black holes ``at
merger''.  NR simulations indicate that the best-fit values for the
coefficients appearing in the above formula are $A = 1.2 \times 10^4$ km/s, $B
= -0.93$ \citep{Gonzalez:2006md}, $H = (6.9 \pm 0.5) \times 10^3$ km/s
\citep{Lousto:2007db}, and $K = (6.0 \pm 0.1) \times 10^4$ km/s
\citep{Campanelli:2007cga}.  The angle $\xi \sim 145^\circ$ for a wide range
of quasi-circular configurations \citep{Lousto:2007db}, while $\Theta_0$
depends on the mass ratio $q$ but not the spins \citep{Lousto:2008dn}.  The
large value of $K$ implies that the merger of equal-mass black holes with
maximal spins pointed in opposite directions in the orbital plane generates a
recoil of $K/16 = 3,750$ km/s.

The most obvious way to avoid these large kicks is to require black
holes to be non-spinning:
\beq
\xI = 0 \to \VD = 0 \to v_\perp = v_\parallel = 0~.
\eeq
In this case, $v_m$ is maximized at the modest value $175 \pm 11$ km/s
for a mass ratio $q \simeq 0.36$ \citep{Gonzalez:2006md}.  However,
theory shows that non-spinning black holes can be spun up to the Kerr
limit $|\boldsymbol{\chi}| = 1$ by steady accretion after increasing
their mass by only a factor of $\sqrt{6}$ \citep{Bardeen:1970}.
Observations of Fe K$\alpha$ fluorescence can be used to measure
black-hole spins \citep{Reynolds:1998ie}, and indicate that real SBHs
can approach this limit: for the SBH in the Seyfert galaxy MCG-06-30-15
the measured spin is $|\boldsymbol{\chi}| = 0.989_{-0.002}^{+0.009}$ at
90\% confidence \citep{Brenneman:2006hw,Berti:2009}.

If black holes are highly spinning, the recoil can be reduced by
aligning the spins and thus $\VD$ with the orbital angular momentum:
\beq
\xI \parallel \LN \to \VD^\perp = 0 \to v_\parallel = 0~.
\eeq
This spin configuration leads to smaller kicks, since the coefficient
$H$ is almost an order of magnitude less than $K$.  The merger of
equal-mass, maximally spinning black holes with one spin aligned with
$\LN$ and the other anti-aligned generates a recoil of $H/16 = 431$
km/s.  Gaseous accretion disks are needed to provide dynamical
friction to allow SBHs separated by $r \simeq 1$ pc to merge in less
than a Hubble time \citep{Begelman:1980vb}.  These same accretion
disks can exert torques on the SBHs which align their spins and
orbital angular momentum with that of the disk, thus producing the
desired aligned spin configuration \citep{Bogdanovic:2007hp}.
However, the effectiveness of this alignment mechanism depends on the
highly uncertain nature of the accretion flow near the merging black
holes.  \cite{Dotti:2009vz} find a residual misalignment of $10^\circ$
($30^\circ$) between the black-hole spins and accretion disk depending
on whether the disk is cold (hot).  This misalignment could be even
greater in a gas-poor merger or one in which accretion onto the SBHs
proceeds through a series of small-scale, randomly oriented events
\citep{King:2007nx,Berti:2008}.

In this {\it Letter}, we present a new mechanism to reduce gravitational
recoils by aligning black-hole spins {\it with each other} prior to merger.
\cite{Boyle:2007sz,Boyle:2007ru} showed that the symmetries of binary
black-hole systems imply that recoils are only generated by a weighted
difference of the two spins.  This general result can be seen to hold
for the fitting formula of \cite{Campanelli:2007ew} by noting that a
weighted difference of spins appears in the numerator of $\VD$ in
equation (\ref{E:Delta}).  Spin alignment is a consequence of
relativistic spin precession as the black holes inspiral due to the
loss of energy and angular momentum to gravitational radiation.  We
begin calculating the inspiral at an initial separation $r_i = 500
R_S$ where spin alignment begins for comparable-mass binaries
\citep{Schnittman:2004vq}, and end at a final separation $r_f = 5 R_S$
near where NR simulations typically begin.  Here $R_S = 2GM/c^2$ is
the Schwarzschild radius of a non-spinning black hole of mass $M$.
Relativists use units in which $G = c = 1$, allowing them to measure
distance and time in units of $M$, where $M \equiv m_1 + m_2$ is the
sum of the masses of the merging black holes. We shall do this for the
rest of the paper. The spin alignment discussed in this {\it Letter}
occurs for both gas-rich and gas-poor mergers, as gravitational
radiation (GR) dominates the dynamics even in the presence of gas at
binary separations less than
\beq \label{E:rGR}
r_{\rm GR} \sim 3000 M q^{1/4} \left( \frac{\dot{M}}{1 M_\odot~{\rm yr}^{-1}}
\right)^{-1/4}~,
\eeq
where $\dot{M}$ is the rate of gas infall \citep{Begelman:1980vb}.

We briefly describe the relativistic dynamics leading to spin
alignment in \S~\ref{S:align}; readers interested in further details
can find them in our longer paper on how spin alignment affects the
distributions of black-hole final spins \citep{Kesden:2010yp}.  The
most notable effect of spin alignment is to suppress the recoil
velocity when the spin of the larger black hole is initially partially
aligned with $\LN$.  The magnitude of this suppression for
distributions with different mass ratios and initial spins is
presented in \S~\ref{S:kick}.  Some concluding remarks are provided in
\S~\ref{S:disc}.

\section{Spin Alignment}
\label{S:align}

To understand why black-hole spins align, we must first describe how they
precess at separations $r < r_{\rm GR}$ where the inspiral is determined
predominantly by the loss of energy and angular momentum to gravitational
radiation.  When the orbital speed is much less than the speed of light, the
motion of the black holes, the precession of their spins, and the emission of
gravitational radiation can all be calculated in the post-Newtonian (PN)
limit.  We use the PN equations for precessing binaries first derived in
\cite{Kidder:1995zr}, supplemented with the quadrupole-monopole interaction
considered in \cite{Racine:2008qv}.  Gravitational radiation circularizes
eccentric orbits \citep{Peters:1963ux}, so we restrict our attention to
quasi-circular orbits of slowly decreasing radius.  In the PN limit $r \gg M$,
the orbital period $t_{\rm orb} \propto r^{3/2}$ (Kepler's third law), the
spin precession period $t_p \sim |\Oi|^{-1} \propto r^{5/2}$, and the
radiation reaction time $t_r \sim r/\dot{r} \propto r^4$.  These scalings
imply that
%
%\beq \label{E:time}
$t_{\rm orb} \ll t_p \ll t_r$,
%\eeq
%
allowing several simplifications.  Since $t_{\rm orb} \ll t_p$, the
spins $\Si$ precess according to
\beq \label{E:Sprecess}
\frac{d\Si}{dt} = \Oi \times \Si~,
\eeq
where $\Oi$ are the orbit-averaged spin precession frequencies.  These
depend on the mass ratio $q$, the orbital angular momentum $\LN$, and
the spins $\Sa$ and $\Sb$ as shown in equation (2.2) of
\cite{Kesden:2010yp}.  Since $t_p \ll t_r$, the total angular
momentum $\mathbf{J} = \LN + \Sa + \Sb$ and magnitude $|\LN|$ are
constant on the timescale $t_p$.  This implies that the direction of
$\LN$ evolves as
\beq \label{E:Lprecess}
\frac{d\hLN}{dt} = -\frac{1}{|\LN|} \left( \frac{d\Sa}{dt} +
\frac{d\Sb}{dt} \right)~.
\eeq
The magnitude $|\LN|$ does decrease on the longer timescale $t_r$ as
the orbital frequency $\omega$ increases according to equation (2.6) of
\cite{Kesden:2010yp}.

\begin{figure*}[t!]
\begin{tabular}{cc}
\includegraphics[scale=0.43,clip=true]{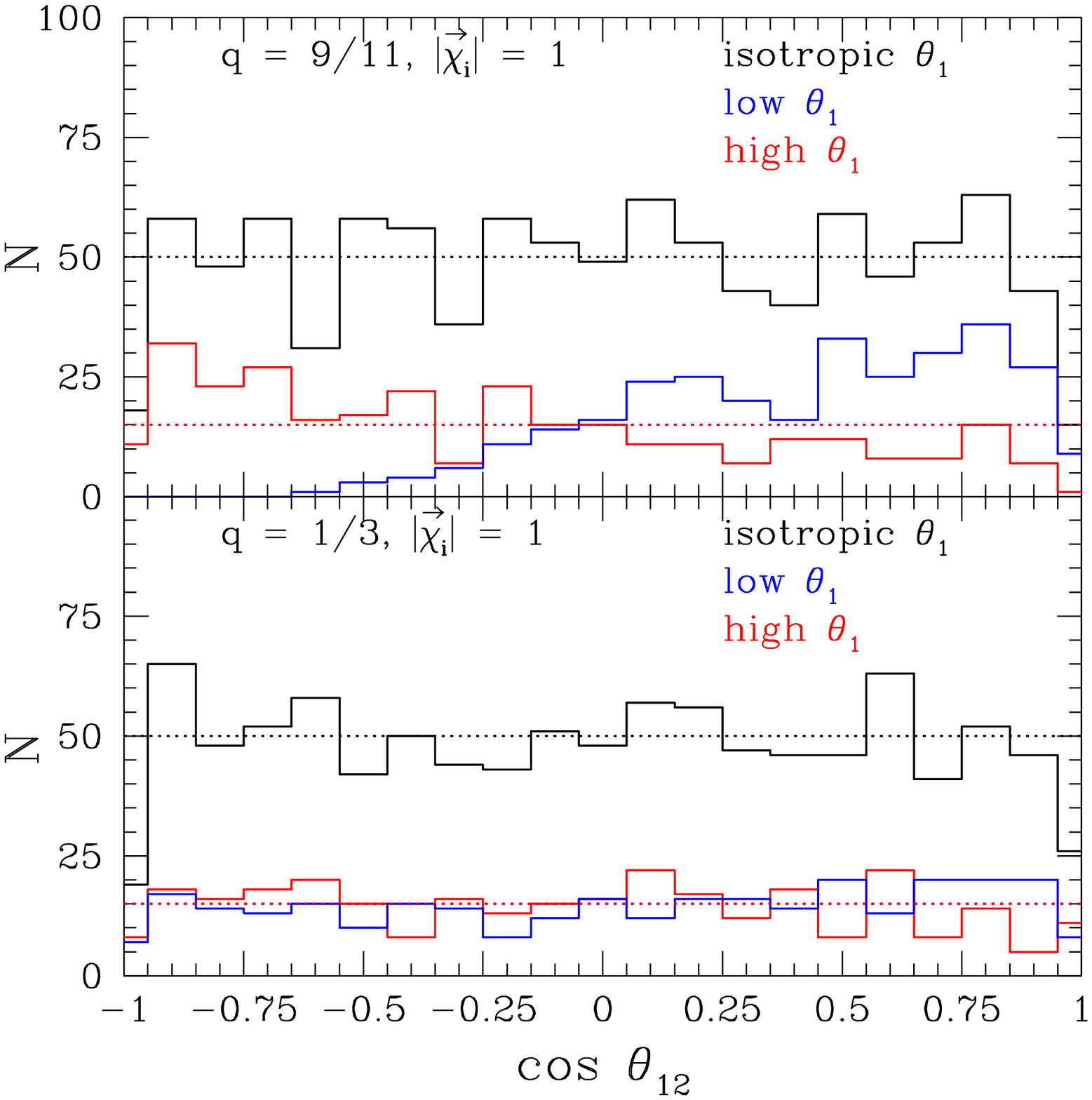}&
\includegraphics[scale=0.43,clip=true]{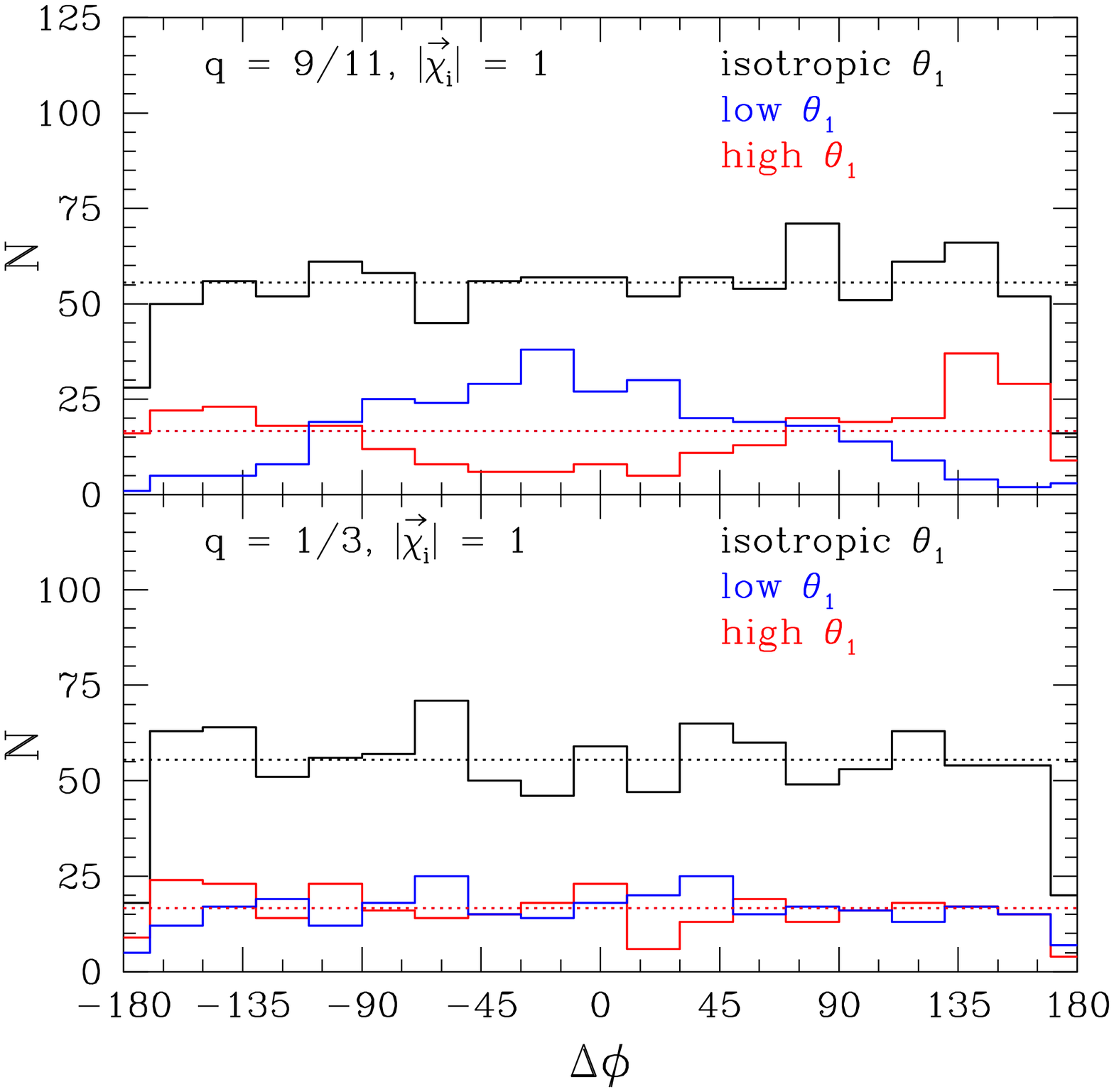}\\
\end{tabular}
\caption{Histograms of $\cos \theta_{12}$ (left) and $\DP$ (right) for
distributions of black-hole spins, where $\theta_{12}$ is the angle
between $\Sa$ and $\Sb$ and $\DP$ is the angle between the projection
of these spins onto the orbital plane.  The horizontal dotted lines
give the distributions at the initial separation $r_i = 1000 M$, while
the solid curves show the distributions after those black holes have
inspiraled to the final separation $r_f = 10 M$.  The black holes are
maximally spinning ($|\xI| = 1$) and have mass ratios $q = 9/11$
(upper panels) or $q = 1/3$ (lower panels).  The black curves
correspond to initially isotropic distributions of the spin $\Sa$ of
the more massive black hole, while the blue (red) curves show
subsets of this distribution with the 30\% lowest (highest) initial
values of $\theta_1$, the angle between $\Sa$ and $\LN$.}
\label{F:align1}
\end{figure*}

\begin{figure*}[t!]
\begin{tabular}{cc}
\includegraphics[scale=0.43,clip=true]{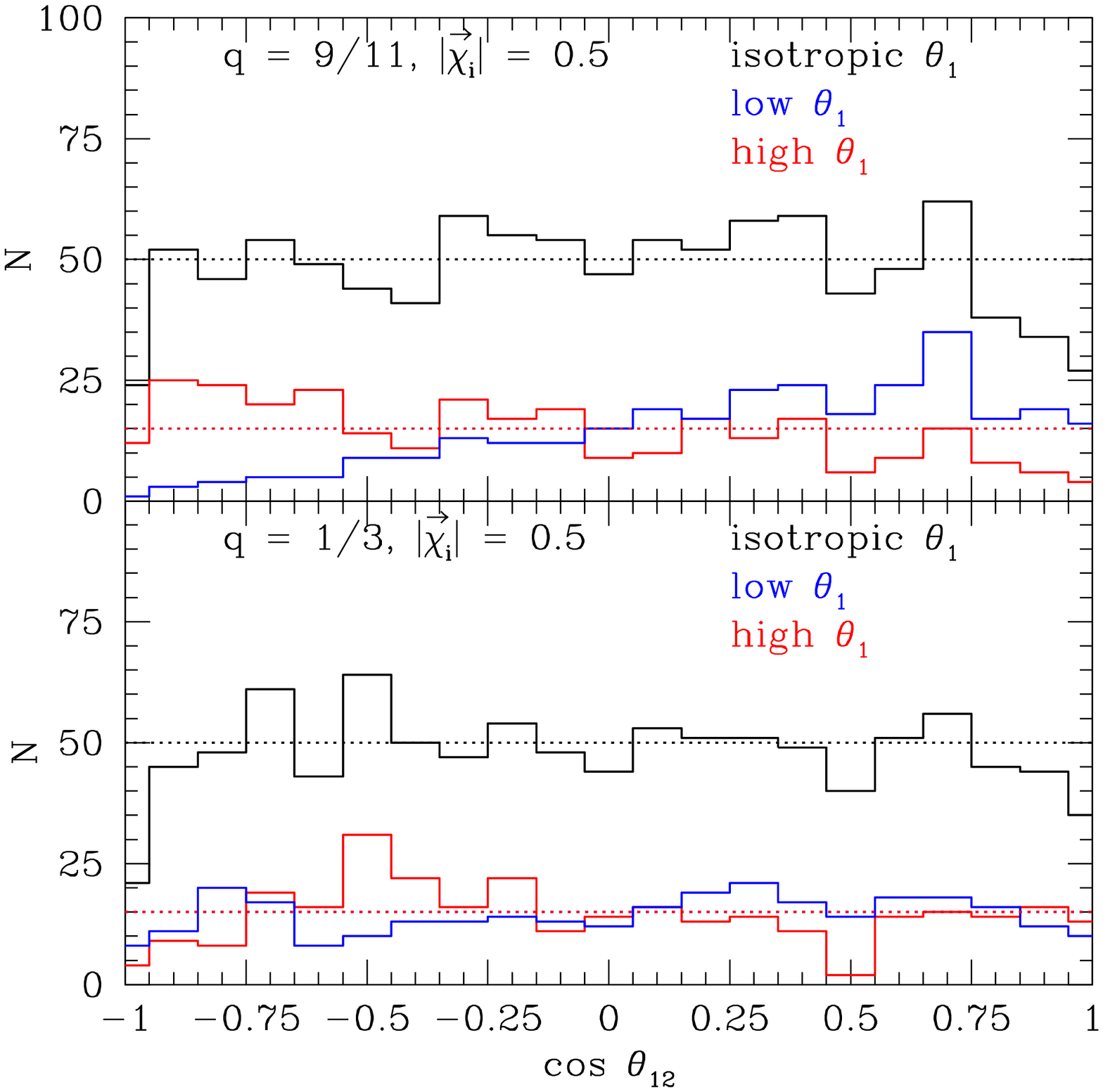}&
\includegraphics[scale=0.43,clip=true]{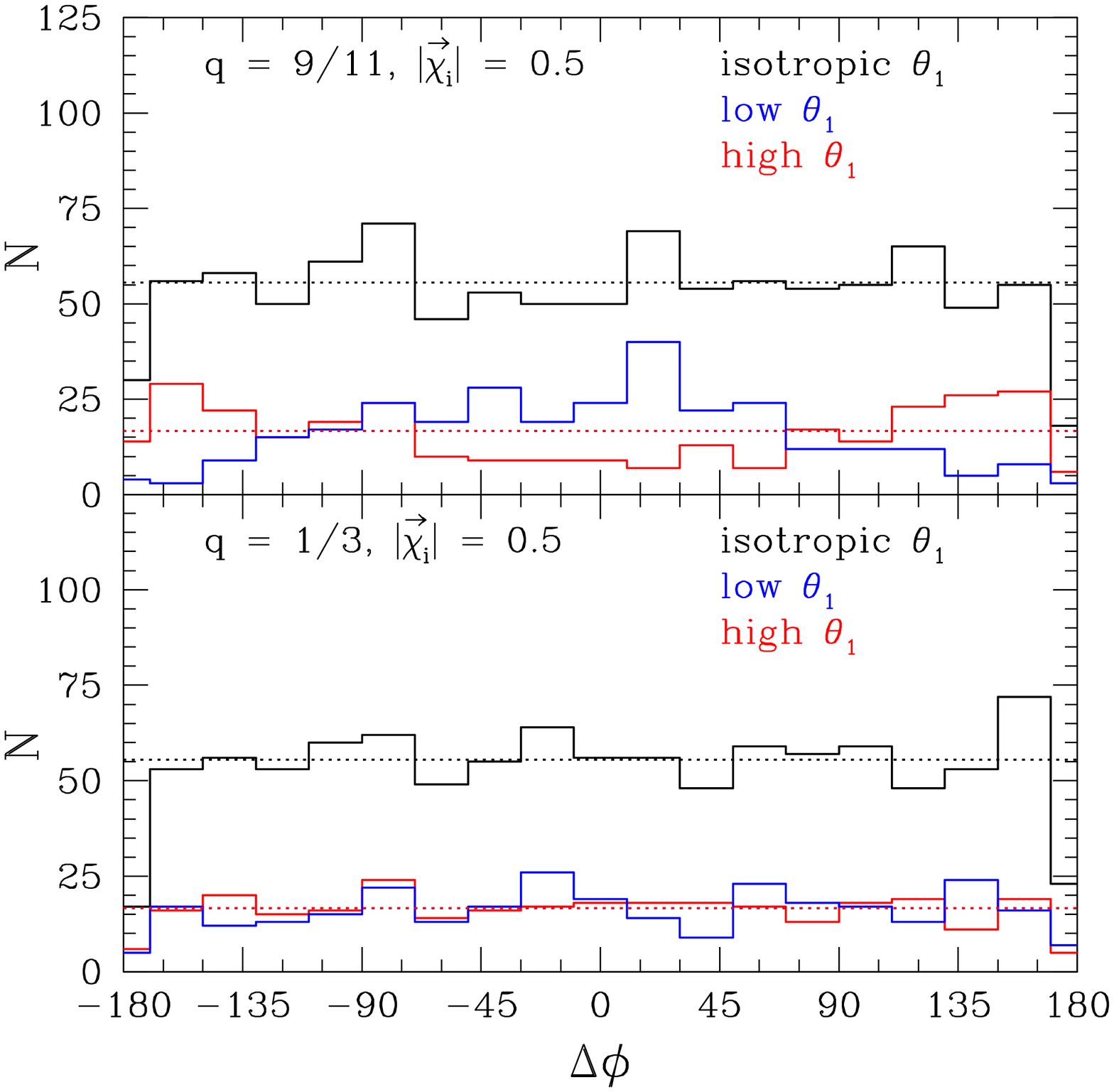}\\
\end{tabular}
\caption{Histograms of the same quantities shown in Fig.~\ref{F:align1},
except the black holes now have initial spin magnitudes $|\xI| =
0.5$.}
\label{F:align0.5}
\end{figure*}

\cite{Schnittman:2004vq} discovered that if black holes inspiral as described
above, spin precession and radiation reaction will act to align $\Sa$ and
$\Sb$ with each other if $\theta_1 < \theta_2$, where $\theta_1$ ($\theta_2$)
is the angles between $\LN$ and $\Sa$ ($\Sb$).  Conversely, if $\theta_1 >
\theta_2$, $\Sa$ and $\Sb$ will become anti-aligned with each other.  This
alignment is strongest for mass ratios $q$ near unity, though it vanishes for
precisely equal masses as there is no distinction between $\Sa$ and $\Sb$ in
that case.  We show the magnitude of this alignment for maximally spinning
black holes ($|\xI| = 1$) in Fig.~\ref{F:align1} and for black holes with
$|\xI| = 0.5$ in Fig.~\ref{F:align0.5}.  The upper panels show black holes
with the nearly equal mass ratio $q = 9/11$, while the lower panels show the
smaller mass ratio $q = 1/3$.  The black curves show that black holes with
isotropic spin distributions at $r_i = 1000 M$ (flat distributions in $\cos
\theta_1$, $\cos \theta_2$, $\DP \equiv \phi_2 - \phi_1$, and $\cos
\theta_{12}$, the angle between $\Sa$ and $\Sb$) maintain these isotropic
distributions as they inspiral to $r_f = 10 M$, consistent with previous
studies of precessing spin distributions
\citep{Bogdanovic:2007hp,Herrmann:2009mr,Lousto:2009ka}.  Isotropic spin
distributions will have $\theta_1 < \theta_2$ just as often as $\theta_1 >
\theta_2$, implying that just as many spins will become aligned as
anti-aligned during the inspiral.

These isotropic spin distributions at $r_i = 1000 M$ are only expected
for the most gas-poor mergers; in the presence of gas, accretion
torques will partially align the spins and orbital angular momentum
with that of the disk \citep{Bogdanovic:2007hp}.  We consider a
scenario in which the spin $\Sa$ of the more massive black hole is
partially aligned with $\LN$ while the other spin $\Sb$ remains
isotropically distributed.  This scenario is consistent with that
explored by \cite{Chang:2009rx} and \cite{Dotti:2009vz}, where the
more massive black hole or ``primary'' is at rest in the center of the
accretion disk while the ``secondary'' migrates inwards.  The blue
(red) curves in Figs.~\ref{F:align1} and \ref{F:align0.5} show the
subset of black-hole binaries with the 30\% lowest (highest) values of
$\theta_1$ at $r_i$.  Since $\Sb$ remains isotropically distributed at
$r_i$, these subsets are consistent with the initially flat
distributions of $\cos \theta_{12}$ and $\DP$ shown by the horizontal
dotted lines.  However, the distributions of $\cos \theta_{12}$ and
$\DP$ no longer remain flat as the black holes inspiral to $r_f$ as
shown by the solid blue and red curves.  Those binaries with low
(high) values of $\theta_1$ at $r_i$ have values of $\theta_{12}$ and
$\DP$ at $r_f$ strongly peaked about $0^\circ$ ($180^\circ$).  This
alignment is very pronounced for $q = 9/11$ both for $|\xI| = 1$ and
$|\xI| = 0.5$, but it is much less significant for $q = 1/3$.

\section{Kick Suppression}
\label{S:kick}

\begin{figure*}[t!]
\begin{tabular}{cc}
\includegraphics[scale=0.43,clip=true]{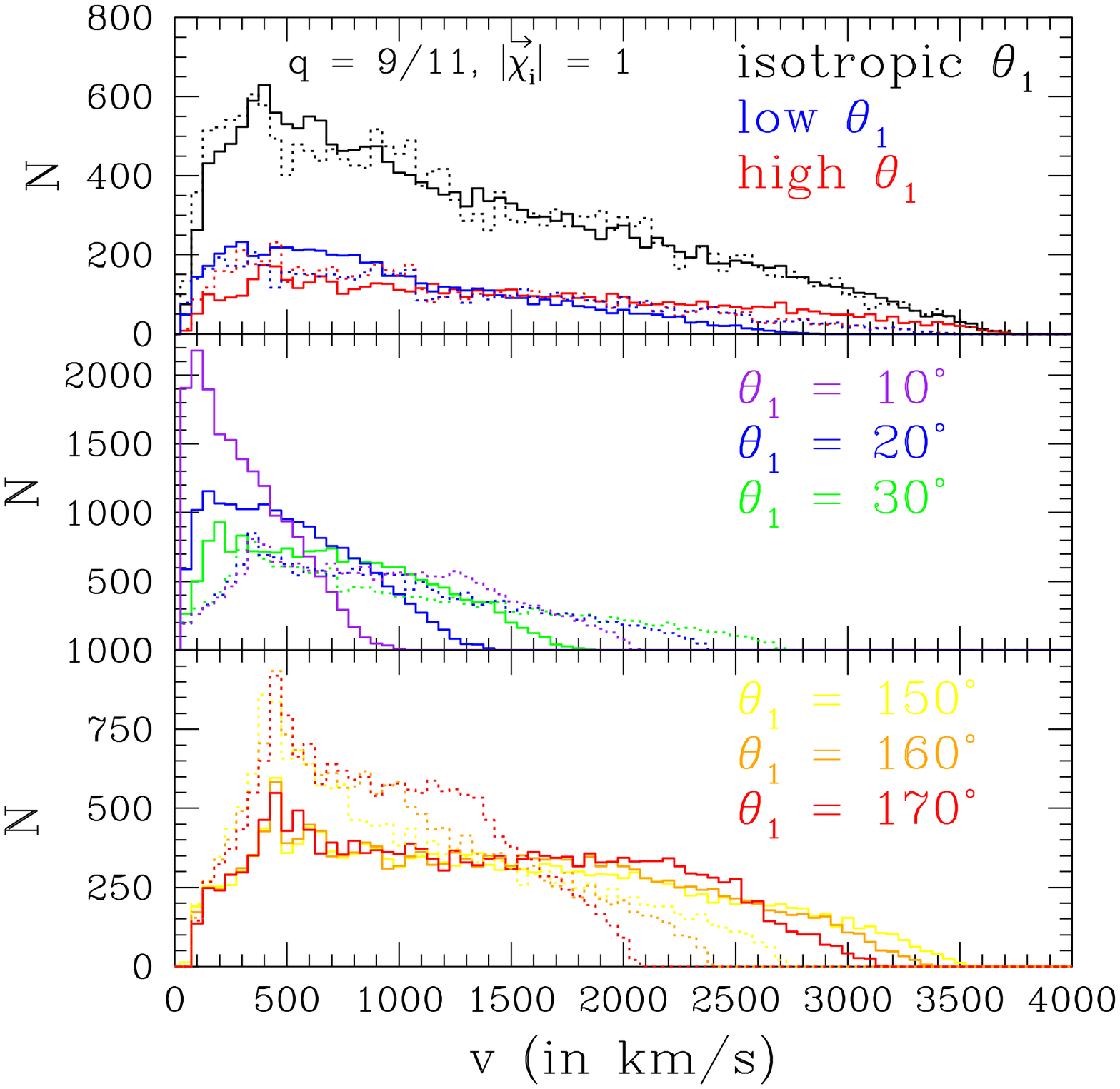}&
\includegraphics[scale=0.43,clip=true]{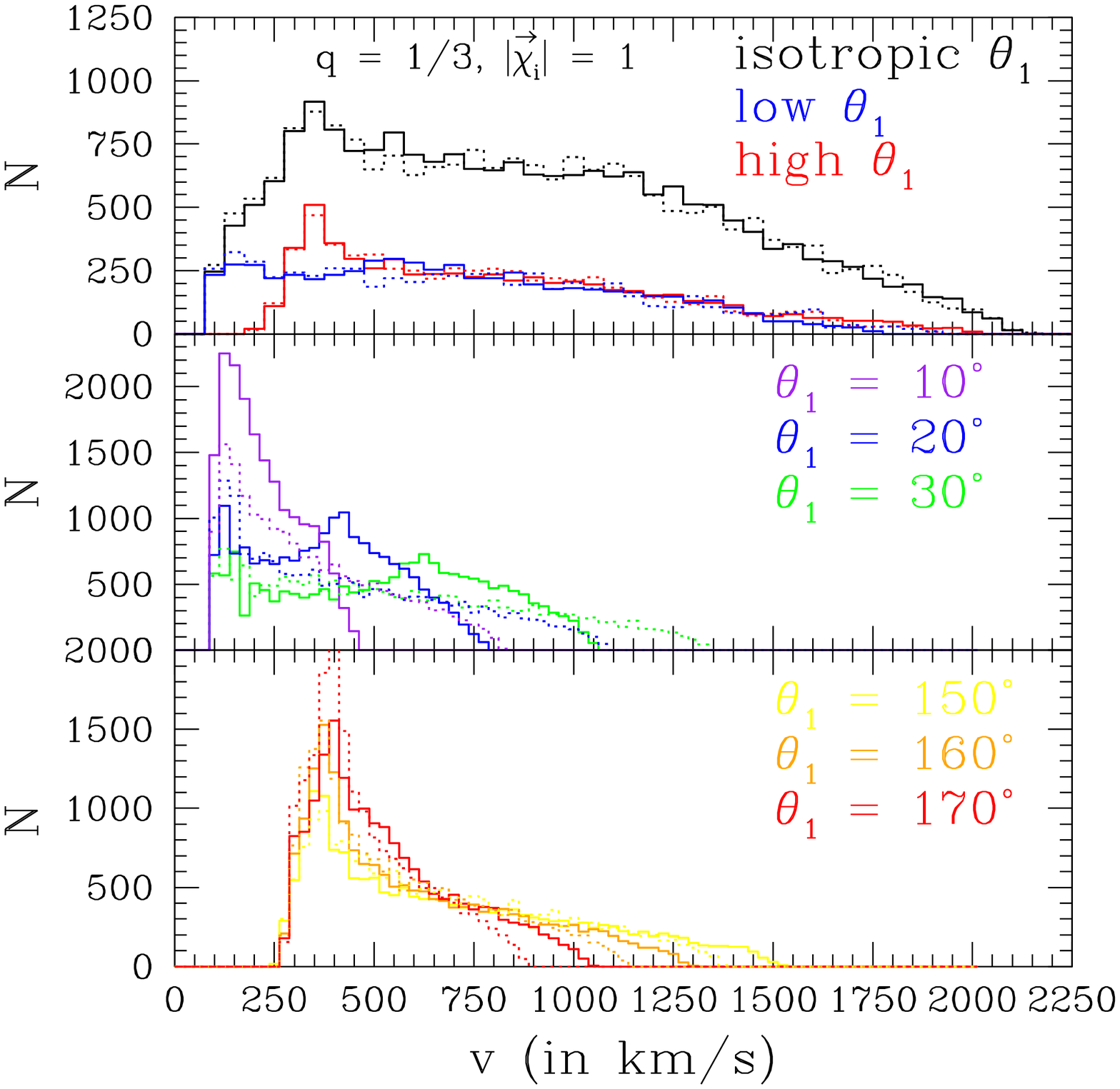}\\
\end{tabular}
\caption{Histograms of the recoil velocity $v$ for maximally spinning
($|\xI| = 1$) black hole mergers with a mass ratio $q = 9/11$ (left)
and $q = 1/3$ (right).  The dotted curves show the recoils expected if
the black holes merge with the spin distributions specified at the
initital separation $r_i = 1000 M$.  The solid curves show the recoils
expected if the spins precess as described in \S~\ref{S:align} as they
inspiral from $r_i$ to the final separation $r_f = 10 M$ prior to
merger.  The black curves in the top panels correspond to isotropic
spin distributions for both black holes, while the blue (red) curves
show the subsets of the spin configurations with the 30\% lowest
(highest) values of $\theta_1$, the angle between the spin of the more
massive black hole and the orbital angular momentum.  The purple,
blue, and green curves in the middle panels show the recoil velocities
expected if initially $\theta_1 = 10^\circ$, $20^\circ$, and
$30^\circ$ respectively and $\Sb$ is isotropically distributed.  The
yellow, orange, and red curves in the bottom panels show the recoil
velocities expected if initially $\theta_1 = 150^\circ$, $160^\circ$,
and $170^\circ$ respectively and $\Sb$ is isotropically distributed.}
\label{F:vhist1}
\end{figure*}

\begin{figure*}[t!]
\begin{tabular}{cc}
\includegraphics[scale=0.43,clip=true]{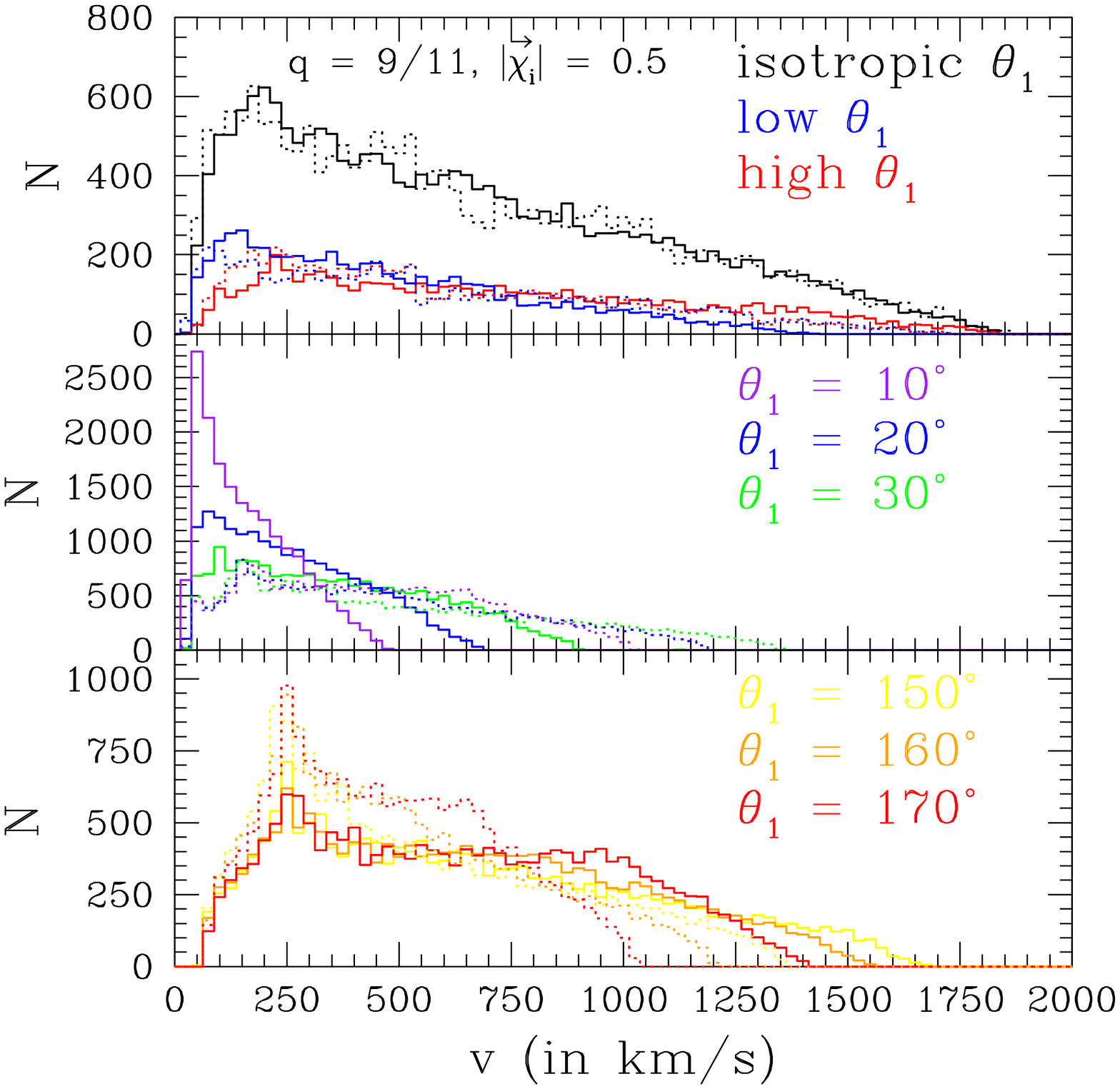}&
\includegraphics[scale=0.43,clip=true]{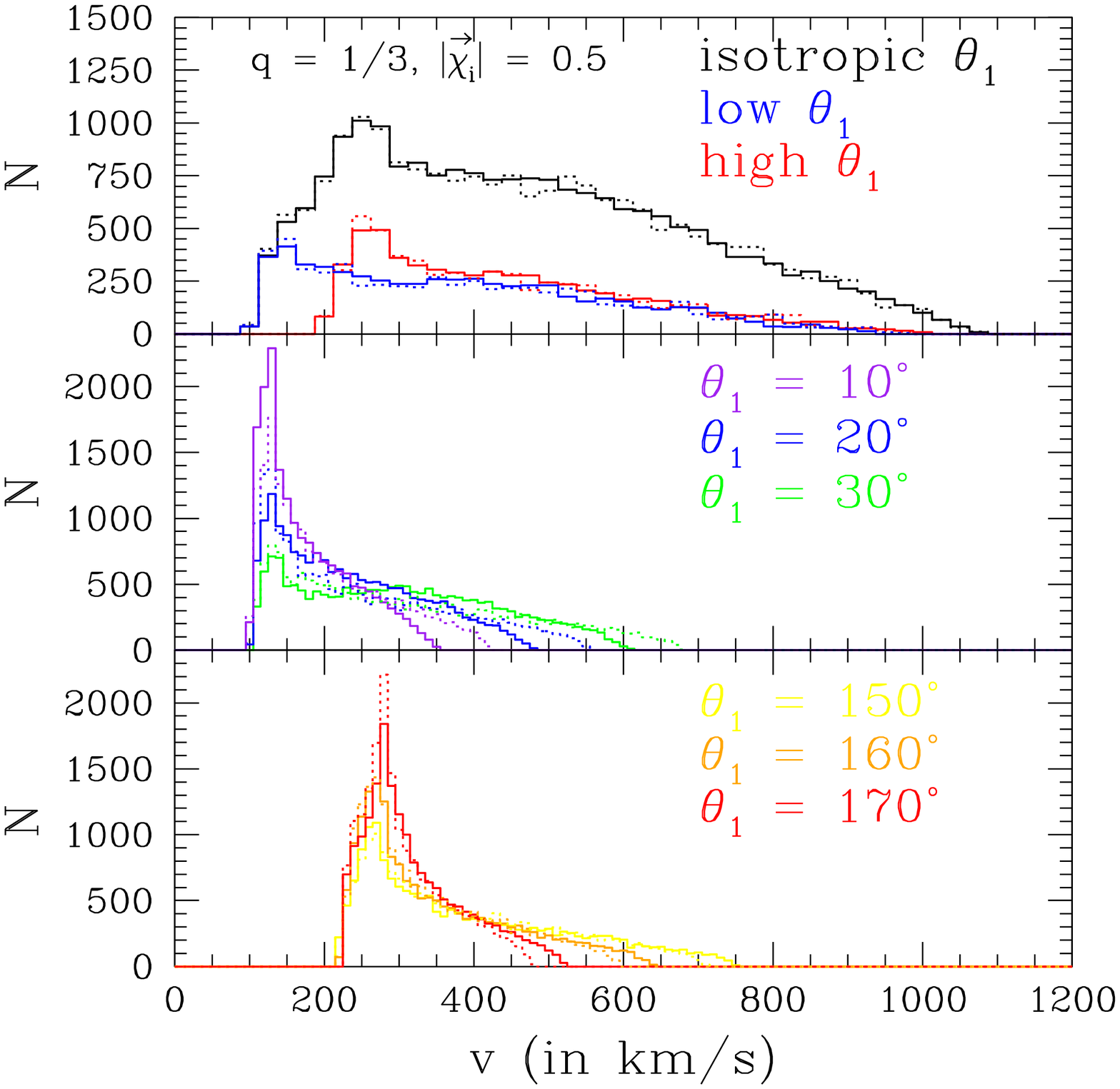}\\
\end{tabular}
\caption{Histograms of the recoil velocity $v$ for black hole mergers
with spins $|\xI| = 0.5$ and mass ratios $q = 9/11$ (left) and $q =
1/3$ (right).  The curves correspond to the same initial spin
directions as described in Fig.~\ref{F:vhist1}.}
\label{F:vhist0.5}
\end{figure*}

How does this alignment of $\Sa$ and $\Sb$ during the inspiral affect
the subsequent recoils?  We show histograms of expected recoil
distributions for maximally spinning ($|\xI| = 1$) mergers with mass
ratios $q = 9/11$ and 1/3 in Fig.~\ref{F:vhist1}, and for mergers with
these same mass ratios and spin magnitudes $|\xI| = 0.5$ in
Fig.~\ref{F:vhist0.5}.  The upper panels of these figures show the
recoil distributions for the same binaries whose spin alignment was
shown in Figs.~\ref{F:align1} and \ref{F:align0.5}.  The dotted curves
in Figs.~\ref{F:vhist1} and Fig.~\ref{F:vhist0.5} show the recoils
expected if the black holes merged with the same spin distribution
they had at $r_i$ (no spin precession), while the solid curves show
the different recoils expected if we include the spin alignment that
occurs as the black holes inspiral from $r_i$ to $r_f$.  Comparing the
dotted and solid blue curves, we see that spin alignment suppresses
the recoils expected if accretion torques at separations $r > r_i$
have partially aligned $\Sa$ with $\LN$ (low $\theta_1$).  The red
curves show that recoils are boosted if $\Sa$ is initially
anti-aligned with $\LN$, but we expect this case to be less physically
relevant.

Spin alignment and the subsequent suppression of recoils can be even
more effective if initially $\theta_1 \leq 30^\circ$ as suggested by
the simulations of \cite{Dotti:2009vz}.  In the middle panels of
Figs.~\ref{F:vhist1} and Fig.~\ref{F:vhist0.5}, we show with purple,
blue, and green curves the expected recoils when $\theta_1 = 10^\circ,
20^\circ$ and $30^\circ$ at $r_i$ while $\Sb$ remains isotropically
distributed.  The dotted curves show that even with this initial
alignment between $\Sa$ and $\LN$, the distribution has tails that
extend to very large recoils.  In Table~\ref{T:CDF}, we show the
velocities $v_{50}~(v_{90})$ at which the cumulative distribution
function of the recoils rises above 50\% (90\%) for various mass
ratios and spins.  The first two entries in the first column of this
table indicate that even if $\theta_1 = 10^\circ$, half the recoils
are greater than 864 km/s and 10\% are greater than 1,587 km/s for $q
= 9/11$ and $|\xI| = 1$.  However, if this spin distribution is
allowed to precess as the black holes inspiral from $r_i = 1000 M$ to
$r_f = 10 M$, the resulting recoils are dramatically suppressed.  The
second column shows that for the same mass ratio and spins, 50\% of
the recoils are below 273 km/s and 90\% are below 611 km/s.  This
reduction is a {\it big} effect: most of the black holes produced will
now remain bound to normal-sized galaxies.  Notice that the reduction
in recoil velocities remains significant even for relatively moderate
spins ($|\xI| \simeq 0.5$).  The precession-induced recoil suppression
is less pronounced for lower spins because the spin alignment is
reduced and the recoil $v_m$ due to the mass asymmetry becomes
comparable to the recoils $v_\perp, v_\parallel$ from the spin
asymmetry.  In this regime, ejection from the largest galaxies seems
unlikely anyway: see e.g. Fig.~2 of \cite{Merritt:2004xa}. For
completeness, we show how kicks are enhanced for high initial values
of $\theta_1$ in the bottom panels of Figs.~\ref{F:vhist1} and
Fig.~\ref{F:vhist0.5}, but don't expect such distributions in
astrophysical mergers.

\section{Discussion}
\label{S:disc}

In this {\it Letter}, we have shown that as black holes inspiral, spin
precession aligns their spins with each other for the spin
distributions expected in astrophysical mergers.  This spin alignment
drastically reduces the recoils expected for the black holes produced
in the binary mergers.  Spin alignment is most effective for the
highly spinning, comparable-mass mergers that are predicted to yield
the largest recoils (up to $v_{\rm max} = 3,750$ km/s according to the
fitting formula of \cite{Campanelli:2007ew}).  Aligning the black-hole
spins generically suppresses the recoils
\citep{Boyle:2007sz,Boyle:2007ru}; we found a similar suppression with
the alternative fitting formula of \cite{Baker:2008md}.  This spin
alignment is a purely relativistic effect that will occur for all
black-hole mergers, as gravitational radiation will always dominate
the inspiral for separations $r < r_{\rm GR} \sim 3000 M$.  As long as
torques at $r > r_{\rm GR}$ align $\Sa$ and $\LN$ such that $\theta_1
\leq 30^\circ$, spin alignment during the final inspiral will nearly
eliminate the $v \gtrsim 1,000$ km/s recoils that are so difficult to
reconcile with galaxies keeping their supermassive black holes.  While
there is still great uncertainty about how merging black holes
interact with surrounding gas, the PN spin precession discussed in
this paper is inevitable and results from well established physics.
We therefore believe that spin alignment must be accounted for in
future population studies of merging black holes.

%--------------------------------------------------------

%\vspace{0.25cm}
%\noindent 
\acknowledgements

We are grateful to Vitor Cardoso for helping to test our numerical
implementation of the PN evolution.  We would also like to thank Enrico
Barausse, Manuela Campanelli, Yanbei Chen, Pablo Laguna, Carlos Lousto, Samaya
Nissanke, Evan Ochsner, Sterl Phinney, \'{E}tienne Racine, Luciano Rezzolla
and Manuel Tiglio for useful discussions. This work was supported by grants
from the Sherman Fairchild Foundation to Caltech, by NSF grants
No.~PHY-0601459 (PI: Thorne) and PHY-090003 (TeraGrid) and by FCT - Portugal
through projects PTDC/FIS/098962/2008, PTDC/CTE-AST/098034/2008 and
PTDC/FIS/098032/2008.  M.K. acknowledges support from NASA BEFS grant
NNX07AH06G (PI: Phinney). E.B.'s research was supported by NSF grant
PHY-0900735.  U.S. acknowledges support from NSF grant PHY-0652995.

%--------------------------------------------------------
% To use bibtex
%--------------------------------------------------------
%\bibliographystyle{unsrt}
%\bibliographystyle{hunsrt}
%\bibliography{adssample}

\clearpage


\begin{table*}
\caption{Velocities $v_{50}$ ($v_{90}$) in km/s at which the cumulative
distribution function for black-hole recoils predicted by the RIT
fitting formula \citep{Campanelli:2007ew} rises above 50\% (90\%).
Entries in each row correspond to distributions with the same spin
magnitudes $|\xI|$ and mass ratio $q \equiv m_2/m_1$.  The columns
indicate the angle $\theta_1$ between the spin of the more massive
black hole and the orbital angular momentum at $r_i = 1000 M$, and the
binary separation ($r_i = 1000 M$ or $r_f = 10 M$) at which the spins
were used to predict the recoil.}
\label{T:CDF}
\begin{center}
  \begin{tabular}{c|c|c|rr|rr|rr|rr|rr|rr}
\hline
\hline
     &     &     & \multicolumn{2}{c|}{$\theta_1=10^\circ$} &
\multicolumn{2}{c|}{$\theta_1=20^\circ$} & \multicolumn{2}{c|}{$\theta_1=30^\circ$} &
\multicolumn{2}{c|}{$\theta_1=150^\circ$} & \multicolumn{2}{c|}{$\theta_1=160^\circ$} &
\multicolumn{2}{c}{$\theta_1=170^\circ$} \\
 $|\xI|$ & $q$ & $v$ & $1000 M$ & $10 M$ & $1000 M$ & $10 M$ & $1000 M$ & $10 M$ & $1000 M$ & $10 M$ &
 $1000 M$ & $10 M$ & $1000 M$ & $10 M$ \\
\hline
1.00 & 9/11 & $v_{50}$ &   864  &   273 &   865 &   474 &   889 &   670 &   910 & 1,354 &   890 & 1,355 &   890 & 1,333 \\
1.00 & 9/11 & $v_{90}$ & 1,587  &   611 & 1,802 &   947 & 2,037 & 1,302 & 2,047 & 2,699 & 1,813 & 2,567 & 1,600 & 2,420 \\
\hline
1.00 & 2/3  & $v_{50}$ &   724  &   283 &   724 &   519 &   801 &   707 &   847 & 1,031 &   777 & 1,035 &   781 & 1,060 \\
1.00 & 2/3  & $v_{90}$ & 1,364  &   538 & 1,602 &   892 & 1,854 & 1,252 & 1,874 & 2,257 & 1,627 & 2,039 & 1,394 & 1,930 \\
\hline
1.00 & 1/3  & $v_{50}$ &   290  &   206 &   382 &   384 &   520 &   562 &   601 &   619 &   495 &   521 &   435 &   488 \\
1.00 & 1/3  & $v_{90}$ &   621  &   364 &   834 &   617 & 1,050 &   878 & 1,093 & 1,183 &   891 &   996 &   697 &   810 \\
\hline
\hline
0.75 & 9/11 & $v_{50}$ &   646 &   198 &   647 &   360 &   665 &   516 &   686 &   966 &   671 &   945 &   672 &   997 \\
0.75 & 9/11 & $v_{90}$ & 1,189 &   423 & 1,350 &   693 & 1,527 &   971 & 1,536 & 1,942 & 1,362 & 1,805 & 1,203 & 1,760 \\
\hline
0.75 & 2/3  & $v_{50}$ &   540 &   210 &   540 &   373 &   599 &   535 &   644 &   785 &   594 &   782 &   597 &   798 \\
0.75 & 2/3  & $v_{90}$ & 1,022 &   439 & 1,200 &   711 & 1,390 & 1,002 & 1,410 & 1,682 & 1,226 & 1,525 & 1,052 & 1,372 \\
\hline
0.75 & 1/3  & $v_{50}$ &   226 &   172 &   294 &   284 &   396 &   411 &   474 &   473 &   400 &   413 &   361 &   388 \\
0.75 & 1/3  & $v_{90}$ &   469 &   330 &   629 &   513 &   791 &   695 &   832 &   895 &   685 &   747 &   546 &   617 \\
\hline
\hline
0.50 & 9/11 & $v_{50}$ &   429 &   142 &   429 &   240 &   442 &   342 &   463 &   609 &   453 &   619 &   453 &   618 \\
0.50 & 9/11 & $v_{90}$ &   792 &   318 &   899 &   491 & 1,018 &   681 & 1,027 & 1,260 &   910 & 1,172 &   805 & 1,103 \\
\hline
0.50 & 2/3  & $v_{50}$ &   359 &   183 &   360 &   246 &   400 &   322 &   444 &   506 &   412 &   506 &   416 &   507 \\
0.50 & 2/3  & $v_{90}$ &   681 &   377 &   800 &   538 &   927 &   679 &   947 & 1,083 &   823 &   966 &   711 &   871 \\
\hline
0.50 & 1/3  & $v_{50}$ &   176 &   158 &   217 &   217 &   281 &   294 &   353 &   354 &   310 &   316 &   289 &   301 \\
0.50 & 1/3  & $v_{90}$ &   326 &   270 &   430 &   374 &   536 &   488 &   577 &   598 &   483 &   509 &   397 &   427 \\
\hline
\hline
0.25 & 9/11 & $v_{50}$ &   217 &   123 &   214 &   144 &   221 &   176 &   242 &   287 &   237 &   290 &   238 &   289 \\
0.25 & 9/11 & $v_{90}$ &   396 &   246 &   450 &   318 &   509 &   398 &   518 &   594 &   461 &   544 &   409 &   503 \\
\hline
0.25 & 2/3  & $v_{50}$ &   189 &   148 &   190 &   161 &   210 &   197 &   251 &   266 &   237 &   261 &   239 &   263 \\
0.25 & 2/3  & $v_{90}$ &   345 &   274 &   405 &   342 &   468 &   411 &   487 &   524 &   429 &   471 &   375 &   421 \\
\hline
0.25 & 1/3  & $v_{50}$ &   156 &   154 &   169 &   168 &   192 &   193 &   244 &   245 &   231 &   232 &   225 &   227 \\
0.25 & 1/3  & $v_{90}$ &   207 &   198 &   251 &   241 &   299 &   290 &   334 &   340 &   296 &   301 &   262 &   268 \\\hline\hline
  \end{tabular}
\end{center}
\end{table*}

\end{document}